# About a structure of easily updatable full-text indexes

alexander@veretennikov.ru

We consider strategies to organize easily updatable associative arrays in external memory. These arrays are used for full-text search. We study indexes with different keys: single word form, two word forms, and sequences of word forms. The storage structure depends on the size of the key's data. The results of the experiments are given in the context of the proximity full-text search, which is performed by means of additional indexes.

## 1. Introduction

Associative arrays in external memory are commonly used for full-text searches. Different types of keys can be considered, for example, a word, a basic form of a word, and several words. The value of a key is a list of records. One record is a posting. A posting is a record with two fields (*ID*, *P*). There, *ID* is the identifier of a document, and *P* is the position of the corresponding key in the document text, for example, the ordinal number of the word. Associative arrays of that kind are named inverted files or inverted indexes [1].

We consider a full-text proximity search. We search documents in which the queried words occur near each other. To solve this task, we need to store information about each occurrence of each key in each document in the index. For example, let us consider two documents, "1.txt" and "2.txt". In the first file, the word "world" occurs 10 times. In the second file, the word "world" occurs 5 times. Therefore, for the key "world", we need to store 15 postings in the index.

The associative array consists of the following components.

1) Data file. The data file is used to store lists of postings. For a specific key, its list of postings should be stored mostly sequentially.
2) Dictionary. The dictionary is used to store keys. For each key, the dictionary contains information about where in the data file the key's list of postings is stored.

## 2. Methods of inverted file construction

### 2.1 Index update

Assume we processed a set of documents and constructed an index. After that, we obtained an additional set of documents. Now, we need a new index that contains information about all our documents, the new set and the previous set.

### 2.2 Method 1. External sorting

When we construct an index, we do the following:

1) We read the documents and write all postings in the data file.
2) We sort the data file according to the key. After sorting, the postings that belong to a specific key are stored in sequence.

When we need to update the index, we do the following:

1) We create a new index using the new set of documents.
2) To obtain the final index, we merge the new index and the previous index [1].

If we need to update the index several times, we do not need to merge at each index update. We can maintain several indexes and sometimes merge some of them [2].

## 2.2 Method 2. Easily updatable indexes

We organize the data file as a collection of blocks. The posting list for a specific key is stored in several blocks. These blocks can be located in different parts of the data file. To create an index, we need to write postings in the index. When we need to write a new posting into the index for a specific key, we locate the posting list for that key in the index and append new data into this posting list. During this process, new blocks can be added to the updated posting list storage.

When the index needs to be updated, we update the existing index in the same way as we created index before, i.e., we add new postings into the corresponding posting lists. See [3, 4, and 5] for more detail.

To create this kind of index, we usually need more input/output operations than those needed to create an index using external sorting. To address this problem, different methods of cache organization and different strategies for maintaining lists of blocks can be considered. Data blocks for a specific key should be stored mostly sequentially. It is required to reduce the number of input/output operations when the search is performed. The author proposed several methods to address these problems [5, 6]. The advantage of this approach is that we do not need to perform a merging procedure when we need to update the index.

The first and second index organization methods have some specific advantages and disadvantages. For the first method, the costly merging procedure is required when we need to update the index. For the second method, more input/output operations are required when we construct the index than are needed for the first approach.

The possibility of a fast index update can be important. In [7, 8, 9, 10], the author considered several index organization variants. In these index organizations, we can use several word forms as a key, not only one word or one form of a word. These additional indexes can be used to significantly improve the search speed when a proximity full-text search is needed. In [8], index construction algorithms are considered. In this approach, we divide the entire text collection, which we are indexing, into several large parts. The size of each part is dependent on the amount of available RAM. Usually, the size of each part is approximately 10-20 gigabytes. In the index constriction process, we add the data of each part in the index. We process each part one by one. Therefore, we perform index updates several times. In [7, 10], we show that with additional indexes, we can improve the search speed by several orders of magnitude.

## 3. Structure of easily updatable index

Easily updatable indexes are created using the second method of index organization. In the current paper, we consider several strategies of easily updatable index construction. We present the results of experiments of index construction. We present a new strategy of the organization of a list of blocks. In this strategy, we use a chain of blocks. The length of the chain is limited by some parameters. This strategy can be used for fast index updates.

Our data file is organized as a sequence of blocks. All these blocks have the same size. These blocks are named clusters. We define the size of a cluster before index construction. Usually, we use clusters with a size of 32 kilobytes.

The stream of clusters is the set of connected clusters that contains one posting list. The data for a specific key are stored in a stream of clusters. An example of a stream of clusters is a chain of individual clusters connected together, similar to a list of clusters.

## 4. Chain of individual clusters

Let us consider a key. When the first posting for the key is added into the index, we can allocate a single cluster for that key. The cluster should contain a link (i.e., an ordinal number of a cluster) to the next possible cluster. We reserve the space for the link in the allocated cluster.

When we need to add more postings for this key into the index, we use the selected cluster and append data sequentially in the selected cluster. Eventually, the available space in the cluster will be used. Therefore, we allocate a new cluster. We connect two clusters. We update the link in the former cluster by the number of new cluster. Then, when a new posting needs to be added, we write data into the new cluster.

Let us consider an example. We show three clusters in Fig. 1. Two clusters are used completely. The last cluster is used partially. In the first and the second clusters, we store the link to the next cluster of the chain. We always add new data into the last cluster of the chain. For the specific key, we store in the dictionary the following: the number of the first cluster of the chain and the number of the last cluster of the chain. The small black box at the end of each cluster represents the area that is used to store the link to the next cluster.

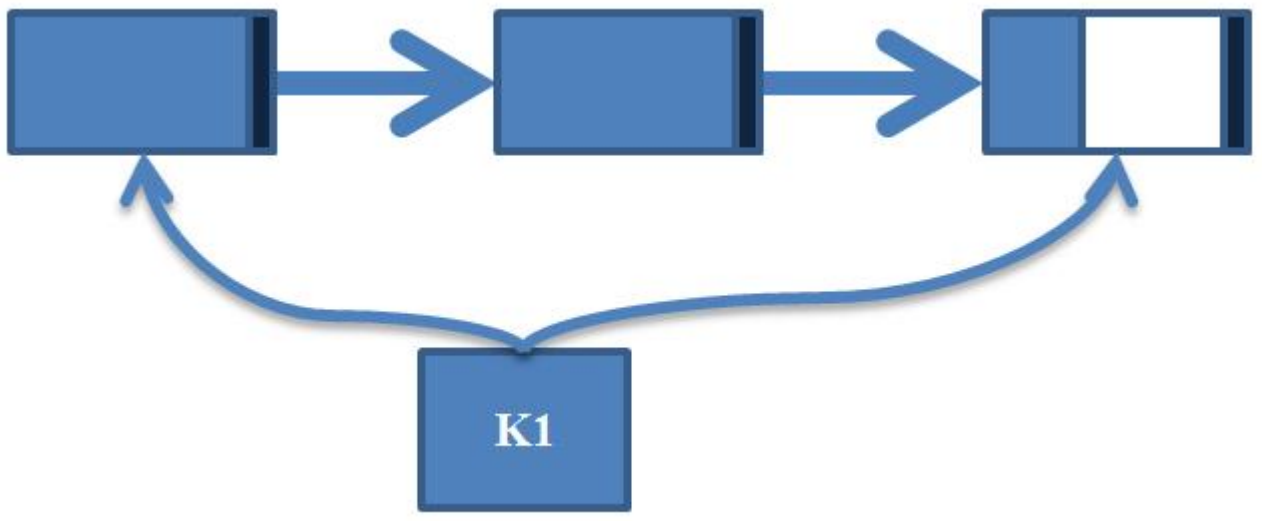

Fig. 1. An example of a chain of individual clusters. For key K1, in the dictionary, we store the numbers of the first and last clusters of the chain.

However, we cannot use only this approach. The number of input/output operations that is needed for the search should be limited. Each cluster of the chain can be located in different areas of the disk. If the length of the chain is large, then we need a large number of input/output operations when the search is performed. Therefore, we cannot always use this approach. In the following sections, we consider several more complex strategies. These strategies allow us to limit the number of input/output operations when the search is performed. Therefore, these strategies provide the possibility of using this kind of index.

# 5. Strategies of easily updatable index organization

## 5.1 Strategy C1

This strategy is related to cache organization. For each stream of clusters, we store at least one cluster in operational memory when the index construction is performed. Let us consider a case in which we use a form of a word as a key. We use a morphological analyser. In the dictionary of the analyser, we have approximately 200 thousand different forms of words if only the Russian language is considered. For each key, the stream of clusters can be created. We can divide the entire set of keys into a set of groups. The size of each group can be, for example, 2000. We can separate the indexing process into phases [5, 8]. Each phase corresponds to a group of keys. In each phase, we only add to the index data for the specific group of keys.

We need to store at least one cluster of each stream of clusters in memory, which is a strong requirement. However, we can store a larger number of clusters per stream of clusters in memory, e.g., 10 or 15. This can improve performance.

The notation C1 is produced from the words “cluster” and 1.

## 5.2 Strategy EM

Let us consider a key. If the posting list for the key is short, the data of the posting list can be stored in the dictionary with the key.

The notation C1 is produced from the word “embedded”.

## 5.3 Strategy PART

Let us consider a key. If the size of the data that is needed to store the posting list for the key is less than half of the cluster size, then only part of a cluster can be used for this key.

We divide a cluster into equal parts. Each part of the cluster can be used for a different key. We use clusters that are divided into different numbers of parts. For example, we use clusters that are divided into two parts, clusters that are divided into four parts, eight parts and so on.

Let us consider a key. When the size of the data for the key is small, we can use a cluster, which is divided into the maximum number of parts. The size of each part of the cluster is small. When the space in the part of the cluster, which we use for the key, becomes insufficient, we move data from this cluster into a cluster with larger parts (and, consequently, the new cluster contains fewer parts).

When the size of the data for the key becomes more than half of the cluster size, we move the data into a single individual cluster. See in more detail in [5].

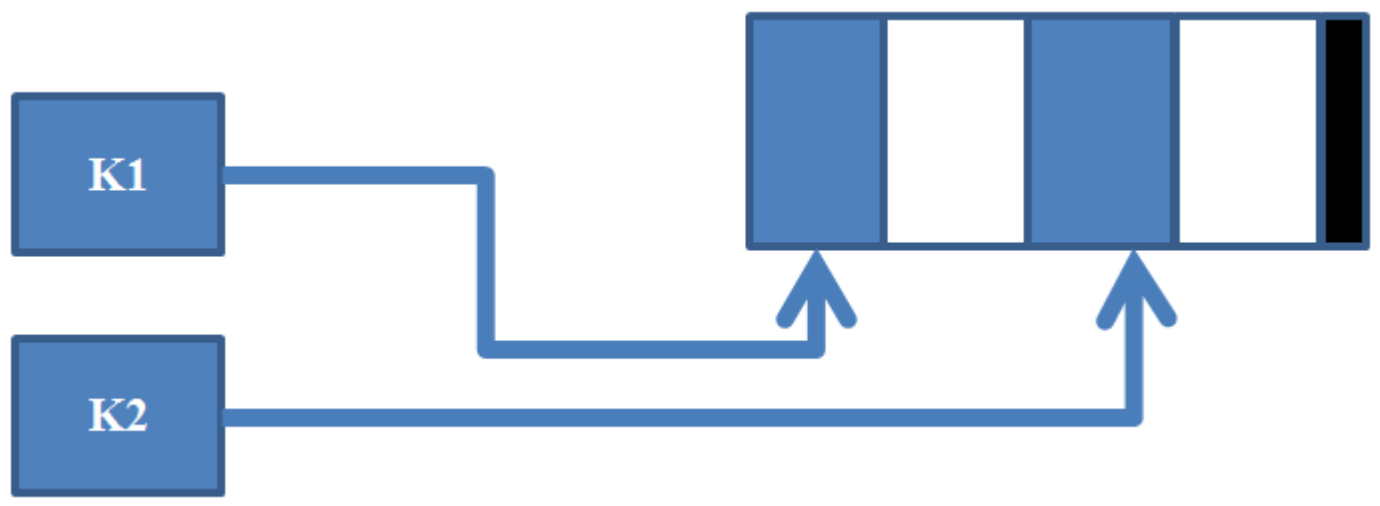

Fig. 2. A cluster that is divided into parts.

Let us consider Fig. 2. Each of the two keys uses only one part of the cluster. The cluster is divided into four parts. Two parts are in use. Two parts are free. The small black box at the end of the cluster represents the metadata area. In the metadata area, we store the information about which parts of the cluster are used. This kind of cluster we call PART-clusters.

The notation PART is produced from the word "part".

## 5.4 Strategy S

Let us consider a key. If the size of the data for the key is greater than the size of one cluster, we need to use several clusters for this key. We introduce the following notation: "segment" of clusters. The segment of clusters is several clusters that are located on the storage device one after another, in sequence. Our segments have lengths that are powers of 2.

When it is required, we allocate a segment of clusters for the key. We add new data into the segment in sequence. Eventually, the available space in the segment is used. In this case, we allocate a new segment. The size of the new segment is the doubled size of the current segment. Then, we move the data from the current segment into the new segment in its first half. We introduce a new parameter $N$. We limit the maximum length of a segment by $N$.

If all available space in the current segment is used and the segment has a maximum length, then we allocate a new segment with maximum length. Then, we link two segments together. See in more detail in [5]. If we use several segments for the key, each of them should have a maximum size.

Let us consider an example. For example, let $N$ be 8. In Fig. 3, we show a stream of clusters that consists of three segments. The first and second segments are used completely. In the last segment, two first clusters are used completely, and the third cluster is used partially.

The last clusters of the first and second segments contain the link to the next segment for the specific segment.

Usually, the maximum size of the segment is approximately several tens of megabytes (several hundred clusters and more).

The notation S is produced from the word "segment".

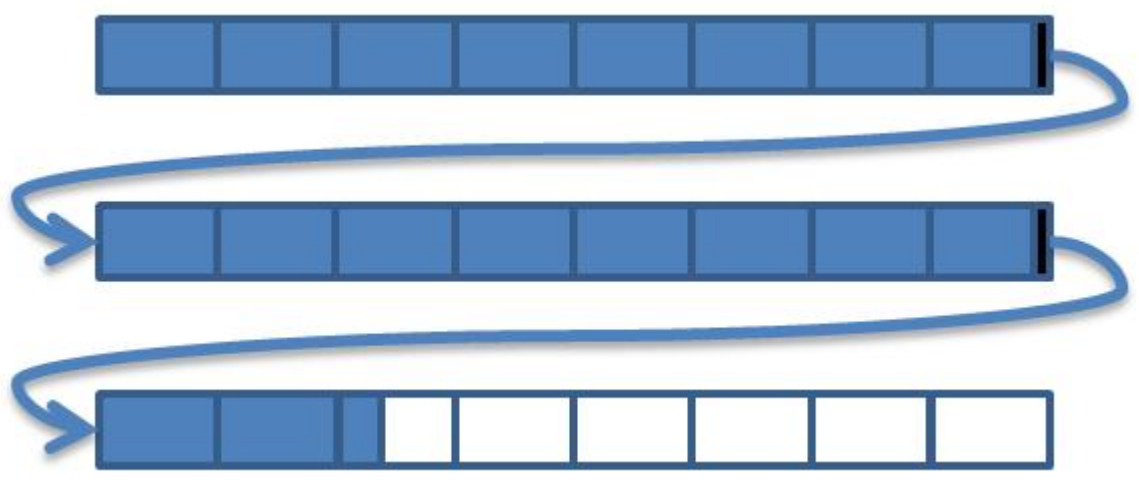

Fig. 3. Segments of clusters.

## 5.5 Strategy FL

In this strategy, we organize a special area of sequentially located clusters. For a specific key, one cluster can be allocated in this area. This cluster is named the FL-cluster. New postings for the key are added into this cluster.

This strategy is used together with some other strategy; for example, with S. Eventually, the available space in the FL-cluster is used. In this case, we move the data from the FL-cluster into the last segment. Then, the FL-cluster can be reused again for new data.

The advantages of this strategy are as follows. When we need to start an update of the index, we can easily read all FL-clusters into our cache. Therefore, for each stream of clusters, we quickly load the last cluster of the stream into the operational memory.

In Fig. 4, we show a stream of clusters that contain two segments. The first segment is used completely. In the second segment, the first five first clusters are used. We write new postings into the separated FL-cluster. The FL-cluster is used partially for now. For key K1, we store the following in the dictionary:

1) The number of the first cluster of the stream.
2) The number of the last cluster used in the last segment of the stream.
3) The number of the FL-cluster.

Let us consider the strategy PART. When this strategy is active, the stream of clusters contains only one cluster. This cluster can also be stored in the FL-cluster area. See in more detail in [11].

The notation FL is produced from the words “first” and “level”. We can say that we divide all clusters into two areas (levels). The first-level clusters are updated when new postings are written into the index. The second-level clusters are more static. Eventually, data are moved from first-level clusters to second-level clusters.

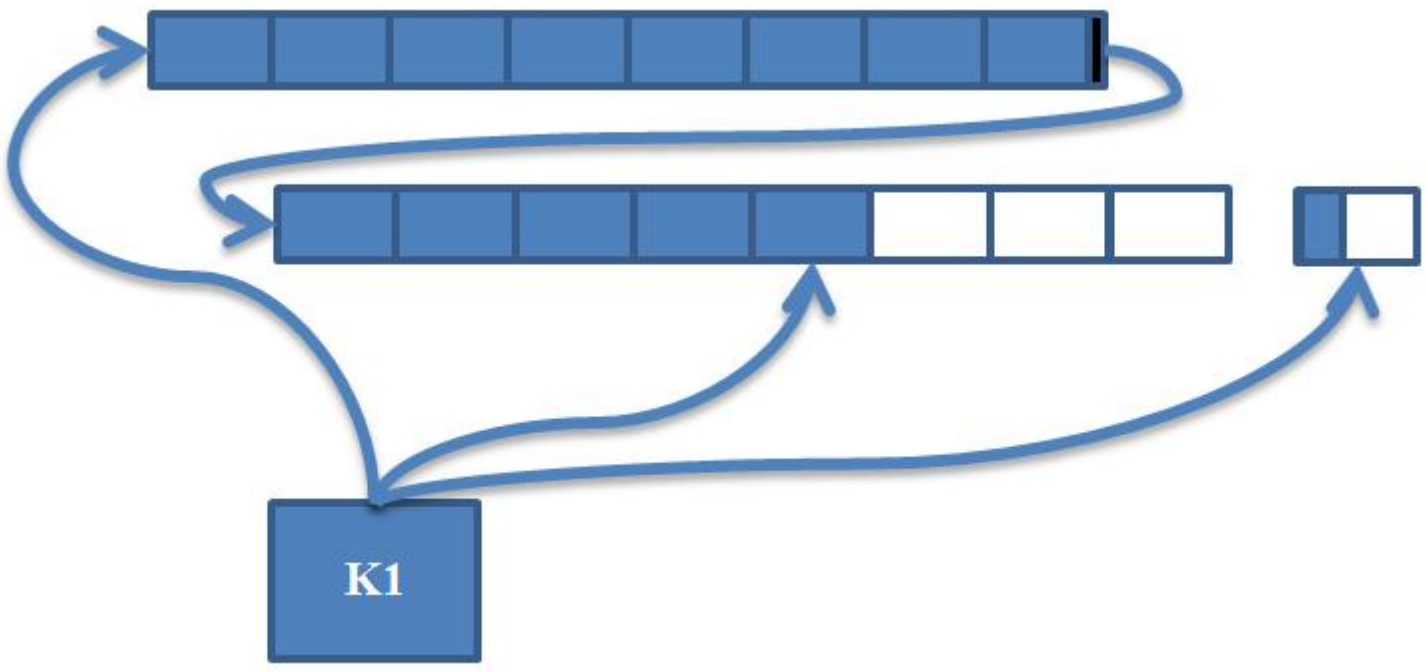

Fig. 4. Two segments of clusters with an additional FL-cluster.

## 5.6 Strategy TAG

In this strategy, one stream of clusters can be used to store data for several keys. To achieve this, every posting contains an additional field. The name of this field is “tag”. Tag contains the local identifier of the key with which the posting is related.

For example, let us have two keys.

For the K1 key, we have the following posting list:

(D1, P1), (D2, P2), . . .

For the K2 key we have the following posting list:

(F1, Q1), (F2, Q2), . . .

We can produce the following combined posting list:

(D1, P1, 1), (F1, Q1, 2), (F2, Q2, 2), (D2, P2, 1), . . .

In this posting list, every posting contains the number of the corresponding key, 1 for K1, and 2 for K2. These numbers are defined locally for this posting list. This posting list is stored in one stream of clusters. This posting list is ordered according to the rules, which are used to order any other list of postings. Therefore, two or more keys can use one stream of clusters. For each key, its local number is stored within the dictionary.

We limit the length of such streams of clusters. Eventually, the posting list of a specific key can become too long. In this case, we extract all data for this key from the combined posting list. A new stream of clusters is created to store these data. The new stream of clusters is dedicated to this key. Therefore, the new stream of clusters does not contain tags. See in more detail in [11, 12].

## 5.7 Strategy CH

### 5.7.1 Chain of clusters with backward links and limited length

In some sense, we return to a simple approach. We use a chain of individual clusters. However, we limit the length of the chain. We add new data into the last cluster of the chain. Eventually, the available space in the cluster will be used. Then, we allocate a new cluster and append it to the chain. We store the number of the previous last cluster in the new cluster.

Eventually, the length of the chain becomes greater than some limit. Let us consider an example. We limit the length of every chain by 9. When the length of a chain becomes greater than 9, we convert the chain into a segment of clusters. This means that we move from the CH strategy to the S strategy.

In more detail, we do the following.

1) Allocate a new empty segment of clusters.
2) Read all data from the chain.
3) Write all data into the new segment.
4) Free all clusters of the chain, i.e., the numbers of these clusters are placed into the special “free clusters” list. Therefore, these clusters can be reused in the future.

This strategy is a new strategy for us (it has not been considered before).

Please note that a chain of clusters can be organized in two ways. Namely, forward links or backward links can be used. In the first sections of the current paper, we discuss forward-linked chains. In this section, we discuss backward-linked chains; see Fig. 5.

When backward-linked chains are considered, we need fewer disk writes than in the case of forward-linked chains. However, for backward-linked chains, we need to read the chain from the end to the start. Therefore, when the search is performed, we can access the first posting of the posting list, only after the entire chain is read. This is not a problem when the length of the chain is limited.

The notation CH is produced from the word “chain”.

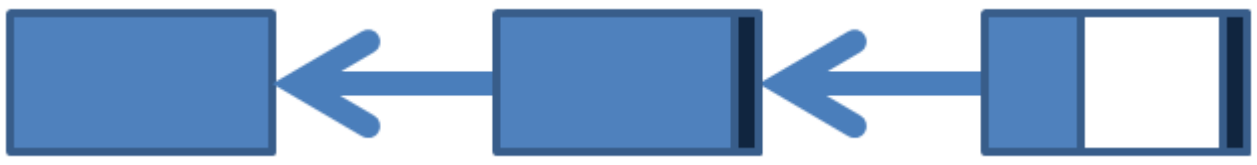

Fig. 5. A backward-linked chain.

### 5.7.2 Chain of segments with backward links and limited length

Assume the following:

1) We use a backward-linked chain.
2) We allocate in the cache several clusters per stream of clusters.
3) We add new data into the backward-linked chain.
4) We see that several clusters at the end of the chain are in cache.

In this case, we can easily move data from these clusters into a new place sequentially, see Fig. 6. Therefore, we will not have a chain of individual clusters but a chain of segments. Eventually, as the size of the data increases, we should move from the CH strategy to the S strategy. The moment of the strategy change depends on the number of segments in the chain (not on the total number of clusters in the chain). This is because the number of segments defines the number of different input/output operations that are required to read the chain's data.

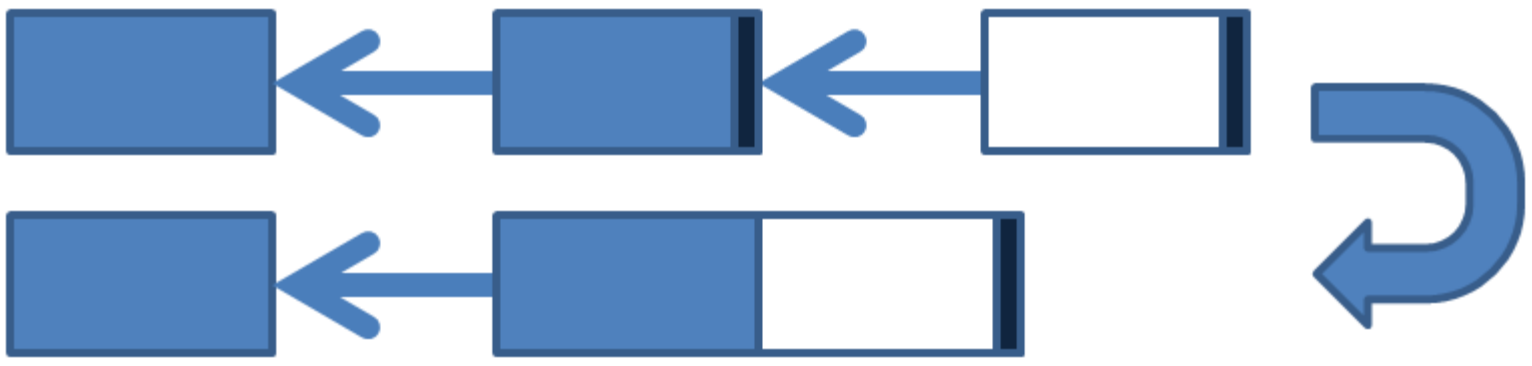

Fig. 6. Conjunction several clusters of a chain of clusters.

The general algorithm of extending the chain, i.e., adding a new cluster to the chain, is as follows. Let us imagine that we added a new cluster to the end of the chain. The purpose of the new cluster is to store some new data. Then, we do the following.

1) We start from the end of the chain. We iterate over segments of the chain, from the end of the chain to the start of the chain. We check segments when moving from one segment to another. If the current segment is not in the cache, then we stop. Therefore, we obtain the list of segments that all reside in the cache.
2) We copy the data from the selected segments to an intermediate buffer (all data reside in the cache; therefore, no disk read operations are required).
3) We allocate a new segment. This segment should be large enough. The size of the new segment should be greater than or equal to the size of the data in the intermediate buffer plus the size of the data that should be stored in the new cluster.
4) We free clusters, which are selected in step 1.
5) We should update the link in the new segment. The new segment should contain the proper link to the previous segment of the chain (this is some segment of the chain, which is not in the cache, on which segment we stopped iteration in step 1).
6) We write new data into the new segment. New data contain the data from the intermediate buffer plus the data from the new cluster. It was established that for the best performance, we should merge at least two last segments of the chain (plus the new cluster). If we always merge only the last segment with the new cluster, then too many free clusters appear (in step 4, we free clusters. That means that we place the numbers of the clusters into some "free clusters" list. These clusters can be reused in the future).

### 5.7.3 The limit of the chain's length

The length of the chain is the number of segments in the chain. The length of the chain should be limited. The limit is related to the total number of read operations, which are required to read the chain's data. For example, we limit the length of the chain by 9. Experiments show that the search time is not changed in comparison with experiments in [10] (where the CH strategy is not used).

In addition, we can set different limits for different streams of clusters. For example, for a specific stream of clusters, we can select a random number from 7 to 9 as the limit of its length. Alternatively, we can use the number of the first cluster of the stream to influence the limit selection. In this case, the transition from the CH strategy to the S strategy would be performed for different streams in different conditions and, therefore, in different index update operations.

## 5.8 Strategy SR

This is a modification of the FL strategy. Instead of using special FL-cluster area in the index, we organize an additional index, namely, the short record index (SRI). A specific stream of clusters can have a record in SRI, namely, SR-record. This record contains a sublist of postings. This record is stored in a list of small blocks. For example, the size of each block can be 128 bytes.

The total size of the SR-record is limited by the cluster size. Let us consider a stream of clusters. Eventually, the size of the SR-record becomes greater than the cluster size. In this case, the data should be moved from SR-record to the stream of clusters.

Accordingly to the C1 strategy, indexing process is divided into phases. At each phase, we process some subset of keys. We add new data for these keys to the index. The SR-records for these keys are stored in operational memory during this process. After the phase is completed, all these SR-records are stored in a special SR-record file. When the new phase is started, all SR-records that correspond to this phase are read from the SR-record file into operational memory. SR-records are stored in the file sequentially. This allows loading or saving SR-records sequentially with buffering.

We also limit the total size of operational memory that is used for SR-records. The SR strategy can only be applied to a stream of clusters if we have available operational memory, accordingly with the specified limit. This means that the SR strategy is applied only for a subset of streams. This strategy is a new strategy for us (it has not been considered before).

The goal of this strategy is to optimize the FL strategy. Consider a stream of clusters. Imagine that the FL strategy is used, and an FL-cluster is allocated for the stream. We cannot predict the utilization of this FL-cluster. The space in the FL-cluster can be used by only 50% or even less. However, we must save the entire FL-cluster on the disk when the index update is completed. Therefore, the total size of the input/output can be significantly greater than the total size of the data. In SRI, we use very small blocks. The sizes of these blocks are significantly less than the cluster size. This allows us to harness this problem.

The SR strategy can be used effectively with the CH strategy. The SR strategy allows us to minimize the number of cluster reads. The goal is to have all clusters in the chain used completely. Let us consider a stream of clusters. Let us imagine that the size of the data that we accumulated in the SR-record becomes greater than the cluster size. Then, we allocate a new cluster in the chain of clusters. We move the data from the SR-record to the new cluster. The new cluster becomes completely used. Some small quantity of data can remain in the SR-record. This allows us to add only clusters that are full to the chain.

Why is it important? Let us imagine that the last cluster of the chain is not completely used. Then, the index update operation is completed. When the next index update is started, eventually, the need to update the chain can occur. In this case, we should read the last cluster of the chain into operational memory because new data should be added into this cluster.

If we add into the chain only clusters that are full, then these extra reads are not required. This is because of the following.

1) The last cluster of the chain is always full. Therefore, new data cannot be added to this cluster. Therefore, we should not read it.
2) We use backward-linked chains. We store the link to the previous cluster in the new cluster when we add a new cluster to the chain. Therefore, we do not need to update the last cluster of the chain when a new cluster is added to the chain.

Let us consider Fig. 7. We show a backward-linked chain here. The SR strategy is also used. The SR-record contains four small blocks. All these small blocks reside in operational memory when the chain is updated. In the dictionary, for key K1, we store the following.

1) The number of the first cluster of the chain.
2) The number of the last cluster of the chain.
3) Reference to the SR-record.

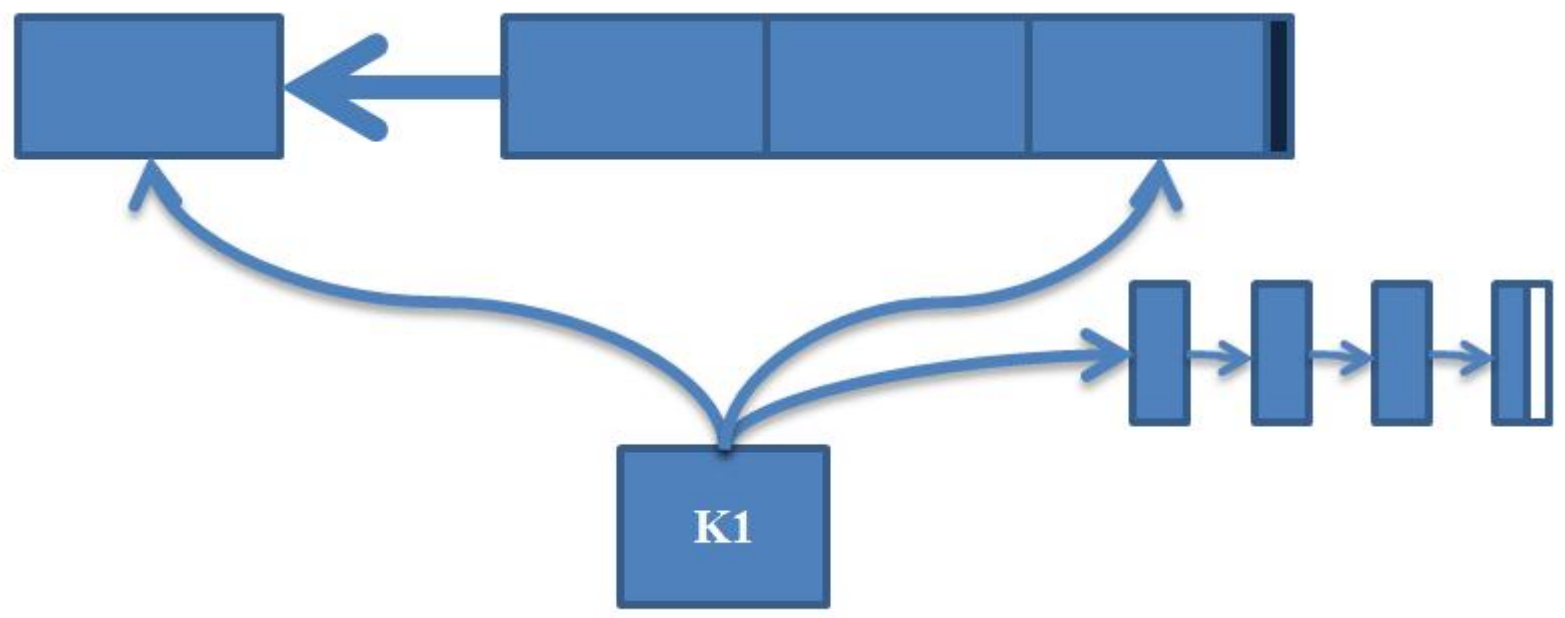

Fig. 7. A backward-linked chain with SR-record.

The notation SR is produced from the words “short” and “record”.

## 5.9 Strategy DS

We use two files. The first file is used for large disk operations. The second file is used for small disk operations. When some number of small disk writes are performed, they are packed into a large buffer. Then, the buffer is stored by one disk write operation. In addition, a mapping table should be used to map the source address of each small disk operation to the target address on the disk, where the data are actually stored. For example, consider three disk writes, with addresses A, B and C. The data are packed into a single buffer. For each source disk operation, we have new addresses for its data, namely, a, b and c, in the buffer. The mapping table should contain the following mapping:

A ⟶ a, B ⟶ b, C ⟶ c. See [9] for more detail.

The notation DS is produced from the words “distributed” and “store”.

## 5.10 A stream of clusters lifecycle diagram

We show in Fig. 8 how the state of a stream of clusters can change over time. The state indicates the strategy that we use. The change in the state depends on the settings and the total size of the data that are stored in the stream.

The TAG strategy is not included in the figure because it is applied independently of other strategies on the dictionary level.

Strategy DS is also organized independently on the file input/output level.

Strategy C1 is always applied.

Let us consider a key. We need to store some data for this key. The data are the posting list. When the size of the data is small, the EM strategy can be used. When the data grow, we should move to the PART strategy or SR strategy. Moreover, when the SR strategy is used and the size of the data is small, we can use only the SR-record to store the data (no clusters are allocated).

Let us consider the PART strategy. The data increase again. Then, the total size of the data becomes more than the size of half of the cluster. We move to the universal S strategy or intermediate CH strategy.

The FL strategy can be used for optimization purposes. For example, we can use the S+FL strategy. This means that the S and FL strategies are used simultaneously, S is the main strategy, and FL is the auxiliary strategy.

Let us assume that the SR strategy is used, the size of the data is small, and no clusters are allocated. The data increase. Eventually, we can move to the S or CH strategy and keep SR as an auxiliary strategy. The PART strategy is not used in this case. Since the SR strategy is used, we allocate an SR-record. The size of the data in the SR-record can already be less than or equal to the cluster size. Therefore, the use of the PART strategy has no sense. The purpose of the PART strategy is to store the data, the size of which is less than half of the cluster size.

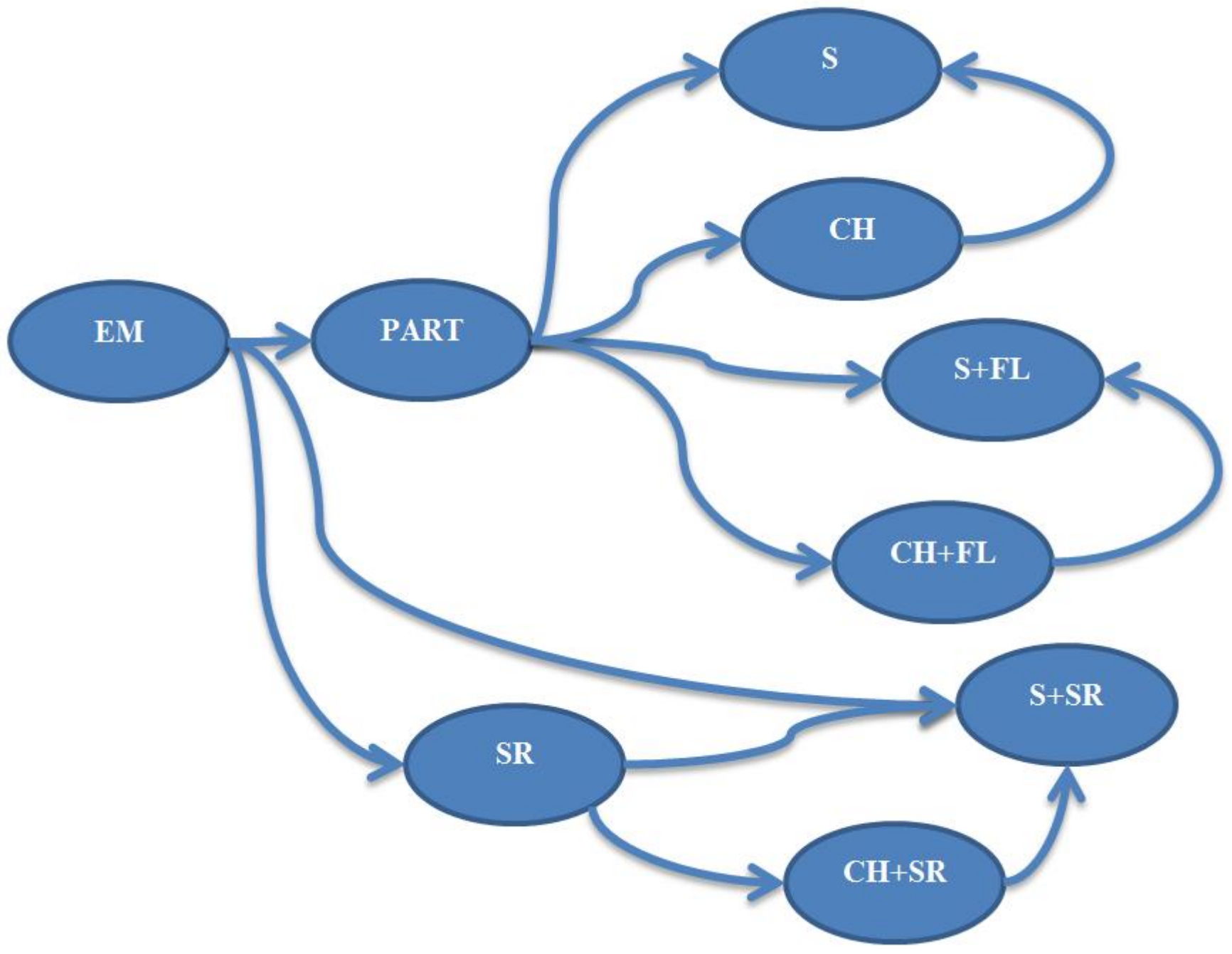

Fig. 8. Diagram of the cluster stream states.

# 6 Using easily updatable indexes for full-text search

## 6.1 Query types

Let us consider the following queries.

Time and a world Yes.

The Who who are you.

End of days.

Tell me, who is your friend.

Words occur in texts at different frequencies. Queries that contain frequently occurring words are complex from the performance point of view. Let us consider a query that contains some frequently occurring words, such as “and” and “who”. Let us construct an ordinary index in which keys are words, and key values are posting lists for the corresponding words. To evaluate this query by means of the ordinary index, we need considerable time.

Let us consider our additional indexes. Each key is a several word forms. With our additional indexes, we can decrease search execution times by several orders of magnitude [7, 10] in contrast with ordinary inverted indexes.

## 6.2 Three types of words

Let us divide all words into three groups.

1) Stop words.
2) Frequently used words.
3) Other.

We used a morphological analyzer. For each word that the analyzer’s dictionary contains, the analyzer provides a list of numbers of base word forms. These words are named *known* words. A base word form number is a number in the range [0, WordCount – 1]. Here, WordCount is the total number of base word forms, approximately 260 thousand for our dictionary.

If we have a word that the dictionary does not contain, then we define that the corresponding base word form is the word itself. These words are named *unknown* words.

Base word forms are also named lemmas. The process of obtaining a list of lemmas for a word is lemmatization. We also divide lemmas into three aforementioned groups.

## 6.3 Three types of indexes

For the following experiments, we used the index structure that was described in [10]. We used the following three types of indexes.

1) Ordinary index. Keys are lemmas.
2) Extended (*w*, *v*) index. Each key (*w*, *v*) is a pair of lemmas. For the (*w*, *v*) key, we store a list of postings in the index. Each posting corresponds to an occurrence of both lemmas *w* and *v* in a text near each other.

3) Index of stop lemma sequences. Each key is a sequence of stop lemmas. For each key, we store a list of postings in the index. Each posting corresponds to an occurrence of the sequence of stop lemmas in a text.

## 6.4 Experiments

We used a text collation that consists of plain text documents with a total size of 71.5 GB. We performed three experiments of index construction. The results of the experiments are presented below. In each experiment, we used a different set of strategies.

1) C1+EM+PART+S+FL+TAG (the first six strategies).
2) All strategies from the first experiment plus CH and SR.
3) All strategies from the second experiment plus DS.

In each case, the total size of the index is approximately 400 GB. In each experiment, we measured the following values.

1) The total size of bytes that were written or read.
2) The total number of input/output operations.

The entire text collection was divided into two parts. We created an index for the first part. Then, we updated the index by indexing the second part.

We show the index parameters in Table 1.

Table 1. The index parameters

| Parameter | Value |
|---|---|
| Size of cluster | 32 KB |
| Number of clusters in the cache per stream of cluster | 45 |
| Threshold value for classification an input/output operation as small (used for DS strategy, [9]) | ≤ 32 KB |
| The total size of the cluster cache | 1 GB |
| The maximum length of each cluster chain (in segments, used for CH) | 9 |
| The number of known lemma groups (we calculate this value based on the total size of cluster cache and the number of clusters in the cache per stream of clusters) [10] | 243 |
| The number of unknown lemma groups | 96 |

All other parameters have the same values as in [10].

## 6.5 Results of the experiments

We show the results of the experiments in Table 2 and Table 3.

The data in Table 2 show that additional strategies help to reduce the total size of input/output. The influence of the DS strategy here is not significant.

The data in Table 3 show that additional strategies help to reduce the total number of input/output operations. The influence of the DS strategy here is significant. With the use of the DS strategy jointly with the CH and SR. We achieve better results in contrast with the selection only base set of strategies.

Table 2. Total sizes of bytes that were written or read (GB)

| Index type | Experiment | | |
|---|---|---|---|
| | 1 | 2 | 3 |
| Known lemmas ordinary index | 209,53 | 190,126 | 192,516 |
| Unknown lemmas ordinary index | 29,571 | 24,16 | 23,774 |
| Extended (*w*, *v*) indexes, *w*, *v* both are known lemmas | 249,555 | 109,731 | 111,04 |
| Extended (*w*, *v*) indexes, *w* is a known lemma, *v* is an unknown lemma | 50,511 | 26,962 | 28,279 |
| Index of stop lemma sequences | 359,023 | 310,645 | 320,552 |

Table 3. Total numbers of input/output operations

| Index type | Experiment | | |
|---|---|---|---|
| | 1 | 2 | 3 |
| Known lemmas ordinary index | 139 694 | 86 115 | 62 993 |
| Unknown lemmas ordinary index | 23 670 | 12 919 | 9 085 |
| Extended (*w*, *v*) indexes, *w*, *v* both are known lemmas | 126 251 | 38 228 | 11 662 |
| Extended (*w*, *v*) indexes, *w* is a known lemma, *v* is an unknown lemma | 50 098 | 33 618 | 26 010 |
| Index of stop lemma sequences | 190 995 | 87 817 | 79 399 |

## 7 Conclusion

In this paper, we introduce several new strategies of easily updatable index organization and construction. These new strategies allow us to reduce the number of input/output operations during the construction of the indexes and, therefore, increase the index construction speed. We apply new strategies to our previously developed algorithms that are used for proximity full-text search. The results of the experiments are presented. We solve the proximity full-text search task by maintaining several additional indexes with different types of keys.

Veretennikov Alexander Borisovich, Candidate of Physics and Mathematics, Associate Professor, Department of Calculation Mathematics and Computer Science, Ural Federal University, pr. Lenina, 51, Yekaterinburg, 620083, Russia. E-mail: alexander@veretennikov.ru

See also:

http://www.veretennikov.ru/

http://www.veretennikov.org/Default.aspx?f=Publish%2fDefault.aspx&language=en